\documentstyle[12pt]{article}
\def\ifundefined#1{\expandafter\ifx\csname#1\endcsname\relax
}
\makeatletter \ifundefined{new@mathgroup} {} \else
    \new@mathgroup\sffam
    \new@internalmathalphabet\mathsf\sffam{cmss}{m}{n}
    \def\psf{\fontfamily\sfdefault \fontseries\default@series
        \fontshape\default@shape\selectfont\mathsf}
\fi
\def\citen#1{\if@filesw \immediate\write \@auxout {\string\citation{#1}}\fi%
\@tempcntb\m@ne \let\@h@ld\relax \def\@citea{}%
\@for \@citeb:=#1\do {\@ifundefined {b@\@citeb}%
    {\@h@ld\@citea\@tempcntb\m@ne{\bf ?}%
    \@warning {Citation `\@citeb ' on page \thepage \space undefined}}%
    {\@tempcnta\@tempcntb \advance\@tempcnta\@ne
    \setbox\z@\hbox\bgroup\ifcat0\csname b@\@citeb \endcsname \relax
    \egroup \@tempcntb\number\csname b@\@citeb \endcsname \relax
   \else \egroup \@tempcntb\m@ne \fi \ifnum\@tempcnta=\@tempcntb
    \ifx\@h@ld\relax \edef \@h@ld{\@citea\csname b@\@citeb\endcsname}%
    \else \edef\@h@ld{\hbox{--}\penalty\@highpenalty
    \csname b@\@citeb\endcsname}\fi
    \else \@h@ld\@citea\csname b@\@citeb \endcsname \let\@h@ld\relax \fi}%
\def\@citea{,\penalty\@highpenalty\hskip.13em plus.13em minus.13em}}\@h@ld}
\def\@citex[#1]#2{\@cite{\citen{#2}}{#1}}%
\def\@cite#1#2{\leavevmode\unskip\ifnum\lastpenalty=\z@\penalty\@highpenalty\fi%
  \ [{\multiply\@highpenalty 3 #1%
  \if@tempswa,\penalty\@highpenalty\ #2\fi}]}   %
\makeatother

\def\cou           {\mbox{$\epsilon$}}
\def\copp          {\mbox{$\Delta'$}}
\def\futnote#1     {\footnote{~#1}\ }
\def\DHR           {Dop\-li\-cher\hy Haag\hy Ro\-berts }
\def\DR            {Dop\-li\-cher\hy Ro\-berts }
\def\hy            {$\mbox{-\hspace{-.66 mm}-}$}
\def\bull          {$\bullet$\,\ }
\def\CGC           {Clebsch\hy Gor\-dan coefficient}
\def\cdt           {\,}
\def\h             {\mbox{$H$}}   \let\H=\h  \def\hh    {\hat H}
\def\id            {{\sl id}}
\def\cop           {\mbox{$\Delta$}}
\def\coa           {\mbox{$\varphi$}}
\def\be            {\begin{equation}}
\long\def\labl#1   {\label{#1}\ee}
\def\apo           {\mbox{$S$}}
\def\bfe           {{\bf1}}   \let\UN=\bfe
\def\coc           {\mbox{$\cal R$}}

\def\O{{\cal O}} \def\A{{\cal A}} \def\cH{{\cal H}} \def\C{{\cal C}}
\def\eps{\varepsilon}

\newcommand{\epij}[1]{\mbox{$e^{i_{#1},j_{#1}}_{p_{#1}}$}}

\def\MS            {{\tt S}} \def\MT            {{\tt T}}
\def\rep           {representation}

\def\stop          {\mbox{$\epsilon$}}
\def\stopt         {\mbox{$\epsilon$}}
  
\def\Co            {{\dl C}}
 \let\dl=\bf  \def\complex       {{\dl C}}
\def\ee            {\end{equation}}
\newcommand{\sumh}[1]{\sum_{#1\in\hat H}}
\newcommand{\sumi}[2]{\sum_{#1=1}^{n_{#2}}}
\newcommand{\sumn}[4]{\sum_{#1=1}^{N_{#2 #3}^{\ \,#4}}}
\newcommand{\F}[8] {\mbox{$F^{(#1)_{\scriptstyle#8}}_{ {#2} #3 {#4},
                   {#5} #6 {#7}}$}}
\newcommand{\Fa}[8]{\mbox{$F^{(#1)_{\scriptstyle#8}}_{ {#4} #3 {#2},
                   {#7} #6 {#5}}$}}
\newcommand{\Fb}[8]{\mbox{$\overline F^{\,(#1)_{\scriptstyle#8}}_{{#2}#3{#4},
                   {#5}#6{#7}}$}}
\def\Fpqr          {\F{pqr}\alpha u\beta\gamma v\delta t}

\newcommand{\FF}[6]{\mbox{$F^{(#1#2#3)_{\scriptstyle#6}}_{#4,#5}$}}

\newcommand{\e}[3] {\mbox{$e^{#2, #3}_{#1}$}}
\newcommand{\cgc}[7]{\mbox{\Large[}\begin{array}{ccc} {}\\[-1.55em]
     \!\!\scs#1&\!\!\!\! \scs#2&\!\!\scs#3\!\! \\[-.32em] \!\!\scs#4
     &\!\!\!\!\scs#5&\!\!\scs#6\!\! \end{array}{\mbox{\Large]}}^{}_{#7}\,}
\newcommand{\cgcs}[7]{\mbox{\Large[}\begin{array}{ccc} {}\\[-1.55em]
     \!\!\scs#1&\!\!\!\! \scs#2&\!\!\scs#3\!\! \\[-1mm] \!\!\scs#4
     &\!\!\!\!\scs#5&\!\!\scs#6\!\! \end{array}{\mbox{\Large]}}^*_{#7}\,}

\def\scs{\scriptstyle}
\newcommand{\erf}[1]{(\ref{#1})}
\newcommand{\R}[5] {\mbox{$R^{(#1 #2)_{\scriptstyle#3}}_{#4, #5}$}}
\let\dstyle=\displaystyle

\def\sigmA   {\sigma}

\def\ii            {{\rm i}}

\let\Bi=\bibitem
  \newcommand{\wb}{\,\linebreak[0]} \def\wB {$\,$\wb}

\begin{document}
\def\thefootnote{\alph{footnote}} \setcounter{footnote}{19}

\rightline{\sf KL-TH-94/14}
\rightline{\sf hep-th/9407013}\vskip 1mm
\rightline{\sf July 1994}\vskip 9mm
\centerline{{\large\bf ON THE QUANTUM SYMMETRY \ \ \,}}\vskip1mm
\centerline{{\large\bf OF RATIONAL FIELD THEORIES}
\raisebox{1.5 mm}{ \futnote{To appear in the Proceedings of the 30th
Karpacz Winter School, Eds.\ J.\ Lukierski et al.} } }
\def\thefootnote{\Alph{footnote}} \setcounter{footnote}{17}
\vskip2em \centerline{\bf
J.\ Fuchs,\,\futnote{{NIKHEF-H}, {Kruislaan 409}, NL -- 1098 SJ~\,Amsterdam,
The Netherlands}\mbox{~}\, \setcounter{footnote}{7}
A.\ Ganchev,\,\futnote{\mbox{{\footnotesize
Humboldt~fellow;~FB~Physik,~Univ.~Kaiserslautern,%
{}~D~--~67653~Kaiserslautern,~Germany}}}\mbox{~}\, \setcounter{footnote}{0}
P.\ Vecserny\'es\,\,\futnote{{Central Research Institute
    for Physics}, {P.O.\ Box 49}, {H -- 1525~\,Budapest}, Hungary }
} \setcounter{footnote}{0} \def\thefootnote{ \arabic{footnote}}

\vskip3em
The aim of this talk is to describe a possible understanding
of the quantum symmetry of two-dimensional ($D=2$)
rational quantum field theories (or $D=1$ chiral rational
conformal field theories). We start by briefly sketching the
operator-algebraic approach to relativistic quantum field
theory (for a review, see for example \cite{HAag,Robe})
and in particular the \DHR\ program for the description of
the superselection sectors ($\!\!$\cite{DHR} for $D>2$ and
\cite{FRS} for $D=2$). The category $\C_\A$ of
localized endomorphisms of the observable algebra is
introduced. This is a strict monoidal, rigid category
which is symmetric in $D>2$, i.e.\ one has permutation statistics,
but only braided in $D=2$, i.e., in $D=2$ one has generically
braid group statistics. We restrict our attention to $D=2$ and to the
rational case, i.e., when $\C_\A$ has a finite number of simple objects.
  In the case of chiral conformal field theories the corresponding
category has been described in \cite{Moore-Seiberg}.
Doplicher and Roberts \cite{DR} have completed the DHR program
in $D>2$, showing that $\C_\A$ is equivalent to the category
of finite-dimensional representations of some compact Lie group
-- the group of ``internal'' symmetries of the theory.

The fact that for $D=2$ the category $\C_\A$ is a braided one has lead
various people to argue that the internal symmetries  are given by
quantum groups.\,\futnote{There is a vast literature on the subject which
we will not refer to -- actually most of\\ \mbox{~~~~~~}\,the works
have been in the framework of rational conformal field theory.}
Considering rational theories one has to restrict oneself to
quantum groups at roots of unity.
For generic values of the deformation parameter
the quantum groups of Drinfeld and Jimbo and
Faddeev\hy Reshetikhin\hy Takhtadzhan
are indeed deformations of the group algebra or
universal  enveloping algebra of ordinary
simple Lie groups or algebras, and in fact the representation
theory remains unchanged. On the other hand, at $q$ a root
of unity much of the similarity with the undeformed case
breaks down. Quantum groups at roots of unity are not
semisimple, and as a consequence their category of representations
contains indecomposable (i.e., reducible, but not fully reducible)
representations.
Though this is well known, a number of  papers simply ignored this fact.
The careful analysis shows that in order to write down quantum group
covariant vertex operators one is forced to include also
the indecomposable representations (for some of the
references see \cite{indec}); these are known only
for the ${\sl su}(2)$ case, and even in this case their analysis
required a large amount of work.

An alternative is to look from the start for quantum symmetries that are
described by a simple Hopf algebra.
The first important step in this direction was made by Mack and Schomerus
\cite{MS} who realized that in order to ``truncate'' the quantum groups to
contain only ``good" representations one has to weaken the
coproduct structure, namely relax the requirement that the coproduct
preserves the unity. This can only be achieved if the
coassociativity of the coproduct is relaxed as well, resulting
in a weak version of Drinfeld's quasi-Hopf algebras.
For related or other approaches see also \cite{rehr,H,M,K,S}.
Here we will restrict  our attention to a certain class of
weak quasi-Hopf algebras \h\ (called {\em rational Hopf algebras})
that serve as the quantum symmetry of rational theories with
fully braided sectors, or equivalently with a maximally extended
observable or chiral algebra \cite{V}.
The important new idea of \cite{V} is to use amplimorphisms \cite{SV1}
instead of representations of \h. This is a very powerful tool
that allows to mimic on a finite-dimensional object (the
rational Hopf algebra)
in a straightforward way most of the information contained
in the infinite-dimensional observable or chiral algebra.
In particular, amplimorphisms possess left inverses,
which leads to conditional expectations, Markov traces,
and a characterization of \h\ by a set of rational numbers --
the statistics weights. Having the right definition
one can              
perform the \DR reconstruction of the
quantum symmetry from the category \cite{SV2,S} (see also \cite{H,K}).

\medskip\noindent
{\bf 1. Haag\hy Kastler nets.}
The algebraic approach to relativistic quantum field theory starts
with a few basic principles -- it is {\em quantum}, so the observables form
a *-algebra of bounded operators; all measurements are
{\em localized\/} in space-time,
so for every bounded (diamond) region $\O$ of space-time there is an
algebra $\A(\O)$ of `measurements performed in $\O$\,';
the theory is {\em relativistic}, so we have Poincar\'e covariance
and, most important, a causal structure (local commutativity), i.e.
$\A(\O)\subseteq \A(\O')'$ where $\A'$ denotes the commutant while
$\O'$ is the causal complement, i.e., points  space-like to $\O$.
The correspondence $\O\mapsto\A(\O)$ is a net ($\{\O\}$
directed by inclusion). The inductive limit is
the quasi-local algebra of observables $\A=\overline{\bigcup_{\O}\A(\O)}$\,.

Of central importance is the vacuum representation $\pi_0$.
By the `electron behind the moon' argument, $\pi_0$ is faithful,
so one can identify $\A$ with $\pi_0(\A)$. We will also
make the assumption of Haag duality in the vacuum
sector, i.e., $\A(\O)=\A(\O')'$.

\smallskip\noindent
{\bf 2. The category of  localized endomorphisms.}
In order to proceed, one has to choose a class of physical
representations describing the `charged excitations' of $\pi_0$.
The DHR criterion selects those $\pi$ that are
localized in a bounded region $\O$, i.e.\ satisfy
$\pi|_{\A(\O')} \simeq \pi_0|_{\A(\O')}$.
Let $V:\cH_\pi\to\cH_0$ be the corresponding unitary and define
$\rho(A)=V\,\pi(A)\,V^{-1}$, $A\in\A$.
{}From Haag duality it is immediate that $\rho$ is a localized
endomorphism of $\A$. Conversely, every localized endomorphism
defines a DHR representation by $\pi = \pi_0\circ\rho$.

The space of intertwiners between $\rho_1$ and $\rho_2$ is
$(\rho_1|\rho_2)=\{T\in\A: T\rho_1(A)=\rho_2(A)T,\, A\in\A\}$.
The localized endomorphisms are the objects and the
intertwiners the arrows of the category $\C_\A$ of
(physical) representations (or of localized endomorphisms) of $\A$.
The gain in introducing endomorphisms is the fact that they can be composed,
and thus we can define the product
of representations, $\pi_1\times\pi_2=\pi_0\circ\rho_1\circ\rho_2$.
If $T_i\in(\rho_i|\rho_i')$ for $i=1,2$, their product is
$T_1\times T_2=T_1\rho_1(T_2)=\rho'_1(T_2)T_1
\in(\rho_1\times\rho_2\mid\rho'_1\times\rho'_2)$.
This product turns $\C_\A$ into a strict monoidal category.
Local commutativity and the transportibility of the endomorphisms
ensure that $\C_\A$ is a braided category. Transporting
two endomorphisms $\rho_i$ by
unitaries $U_i$ to $\tilde\rho_i$ that have space-like separated
supports, one can show that locality implies
$\tilde\rho_1\tilde\rho_2=\tilde\rho_2\tilde\rho_1$.
Then the statistics operator
$\eps(\rho_1,\rho_2)\in(\rho_2\rho_1|\rho_2\rho_1)$ is defined
as $(U_1\times U_2)\, (U_2\times U_1)^{-1}$.
The statistics operator is ``almost'' independent of the transporters
$U_i$ -- for $D>2$ it is completely independent and
$\eps(\rho_1,\rho_2)$ is unique, hence its square is the identity
and braid group statistics reduces to permutation statistics.
For $D=2$ where the space-like complement of a point has
two disconnected components, $\eps(\rho_1,\rho_2)$
depends only the relative left/right positions of
$\rho_i$, and in general $\eps$ is different from its
inverse so that genuine braid groups arise.

Two endomorphisms are equivalent if the corresponding representations are
unitarily equivalent. The set of equivalence
classes $[\rho]$ are  the superselection sectors
of the theory. Irreducible sectors $[\rho_r]$ correspond to
`elementary' particles (the simple objects of $\C_\A$)
 and are characterized by their
`charges' $r$. We restrict ourselves to {\it rational} theories, i.e.,
theories for which the set of simple objects of $\C_\A$ is finite.
In general the product of two irreducible endomorphisms is reducible.
Decomposing it into irreducible components, let
$N^t_{rs}=\dim(\rho_r\rho_s|\rho_t)$. These numbers,
the fusion rules, are independent of the representative, so we can write
$[\rho_r][\rho_s] = \sum_t N^t_{rs} [\rho_t]$\,.

If\, $T^{u,\alpha}_{pq}$, $\alpha=1,\dots, N^u_{pq}$, is a basis
of $(\rho_p\rho_q|\rho_u)$, etc., then we can decompose
$\rho_p\rho_q\rho_r$ in two different ways, thereby obtaining two
bases for  the intertwiner space $(\rho_p\rho_q\rho_r|\rho_s)$.
The change of basis is described by the fusing matrix $F$, i.e., $F$
is defined by $T^{s,\alpha}_{ur} T^{u,\beta}_{pq} = \sum_{v;\gamma,\delta}
F^{pqr,s}_{\alpha u\beta, \gamma v\delta} T^{s,\gamma}_{uv}
\rho_p(T^{p,\delta}_{qr})$. The fusing matrices play a role
completely analogous to $6j$-symbols and in particular
satisfy the pentagon equation (see e.g.\ \cite{Moore-Seiberg}). One has also
a braiding matrix describing the relation between $\rho_p\rho_q$ and
$\rho_q\rho_p$.

Charge conjugation means that there is an involutive map
$\rho\mapsto\overline\rho$ of the sectors, such that
$\dim(\overline\rho\rho|\id)=1$.
Let $R\in(\overline\rho\rho|\id)$, $R^*R=\bfe$. Then one can
introduce the left inverse $\Phi$ of $\rho$,
defined by $\Phi(.) = R^*\overline\rho(.)R$.
Left inverses allow to define Markov traces which give
important numerical characteristics of the sectors.
For example the trace of the statistics operator of
an  irreducible sector $[\rho_r]$ is the
statistics parameter $\lambda_r=\Phi_r(\eps(\rho_r,\rho_r))$,
a complex number.  The statistical dimension is
$d_r=|\lambda_r|^{-1}$, the statistical phase is
$\omega_r=d_r\lambda_r$, and the statistical weight
is $w_r=\arg\omega_r/2\pi\ii$.
The dimension $d_r$ is  the square root of the index of
the inclusion $\rho_r(\A)\subseteq\A$\, \cite L. It can also be characterized
as the Perron\hy Frobenius eigenvalue of the fusion rule matrix
$(N_r)_{pq}=N^q_{rp}$. The statistical weight is the
spin or conformal weight modulo integers.  The  monodromy matrix is defined
as $Y_{pq}=d_pd_q {\rm tr} (\eps(\rho_p,\rho_q)\eps(\rho_q,\rho_p))$
\cite{rehr}. $Y$ is nondegenerate if and only if $|\sigma|=\sum_r d_r^2$,
where $\sigma=\sum_r d^2_r \omega_r^{-1}$, and in this case
the observable or chiral algebra
is in a sense maximally extended (there are no sectors except
the vacuum sector that have trivial monodromy with every other sector).
Moreover one has an action of $SL(2,{\bf Z})$, the (double cover of the)
modular group, on the set of sectors, given by $\MS=|\sigma|^{-1} Y$
and $\MT=(\sigma/|\sigma|)^{1/3}\;{\rm diag}\,(\omega_r)$.

\smallskip\noindent
{\bf 3. Modular fusion algebras (MFA).}
A rational fusion ring is a ring with a finite basis $\{\phi_r\}$
in which the structure constants are nonnegative integers,
i.e.,  $ \phi_r * \phi_s=\sum_t N^t_{rs}\phi_t$
with $N^t_{rs}\in{\bf Z}_{\ge 0}$, which is commutative,
i.e., $N^t_{rs}=N^t_{sr}$, and associative,
i.e., $\sum_u N^u_{pq} N^s_{ur}=\sum_v N^s_{pv} N^v_{qr}$.
Moreover there is a conjugation, i.e., a permutation
$r\mapsto\overline r$ of the labels which squares to the identity
and is an automorphism of the ring,
i.e., $N^{\overline t}_{\bar r\,\bar s}=N^t_{rs}$;
it is implemented by $C_{rs}=N^0_{rs}$
(we denote by 0 the label of the identity element of the ring),
i.e., $\phi_{\overline r} = \sum_s C_{rs}\phi_s$, hence
two conjugate elements fuse into the identity with multiplicity one.

The fusion rules matrices $(N_p)_{qr}=N^r_{pq}$ are normal and commuting,
and hence can be simultaneously diagonalized. If the diagonalization
matrix $S$ can be symmetrized and moreover one can find
a diagonal matrix $T$ such that they generate the modular
group, one says that the fusion algebra is modular \cite F.
According to the remarks above, the sectors of a two-dimensional
rational field theory with maximally extended observable algebra form a MFA.

\smallskip\noindent
{\bf 4. Rational Hopf algebras (RHA).}
The quantum symmetry of a rational relativistic quantum
field theory in $D=2$ is a
rational (i.e., finite-dimensional), semi-simple, quasi-triangular,
weak quasi-Hopf *-algebra with invertible monodromy matrix -- for
short a Rational Hopf Algebra (RHA). Let us explain one by one
the elements in the definition.

\bull
Let $\hat H$ be the finite set of irreducible representations of $H$,
i.e., for every $r\in\hat H$ we have $D_r:H\to M_r=End(V_r)$, where
$V_r={\bf C}^{n_r}$ and $M_r={\it Mat}(n_r\times n_r,{\bf C})$ for some
$n_r\in {\bf N}$.  Hence $H$ is a finite sum of full matrix algebras,
\be
  H = \bigoplus_{r\in\hat H} M_r \,.
\ee
The multiplication in \H\ is the ordinary multiplication of matrices.
The *-ope\-ra\-tion is the usual hermitian conjugation of matrices.

\bull
\H\ is endowed with a coproduct $\cop\colon\ \H\to \H\otimes\H$,
which is a *-mo\-no\-mor\-phism. (The *-operation on $\h\otimes\h$ is
defined by $(a\otimes b)^*=a^*\otimes b^*$.)
It is important to note that in general $\cop$ is not unit preserving,
i.e., $\cop(\bfe)$ is in general only a projector.
The coproduct allows to define products of representations by
$(D_1\times D_2)(a) = (D_1\otimes D_2)(\cop(a))$.
The coproduct is quasi-coassociative. Thus there is coassociator, i.e.\
an element $\coa\in\H^{\otimes 3}$ such that
\be (\cop\otimes\id)\circ\cop(a)\cdot\coa =
  \coa\cdot(\id\otimes\cop)\circ\cop(a)\ \qquad \mbox{for all }a\in\h,
\labl{coa1}
which serves as the natural isomorphism between the
two ways of bracketing a triple  product of representations.
Since in general $\cop(\bfe)\ne\bfe\otimes\bfe$, one cannot ask
for $\coa$ to be unitary, but only to be a partial isometry with
domain $ (\id\otimes\cop)\circ\cop(H)$
and range    $ (\cop\otimes\id)\circ\cop(H)$.

\bull
There is a special one-dimensional representation $\cou\colon\ \H\to\complex$,
called the co-unit (we will denote it also by $D_0$),
which is a unit preserving *-ho\-mo\-mor\-phism, and
there are unitary elements $\rho,\lambda\in\H$ such that
$ (\cou\otimes\id)\circ\cop(a)=\rho a \rho^*$,
 $ (\id\otimes\cou)\circ\cop(a)=\lambda a\lambda^*$\, for all $a\in\h$.
The latter serve as natural isomorphisms between
$D_0\times D_p$, respectively $D_p\times D_0$, and $D_p$.

\bull
For these structures there are also two compatibility constraints.
First, the triangle identity
$  (\id\otimes\cou\otimes\id)(\coa) = (\lambda\otimes\UN)\cdot\cop(\UN)
  \cdot(\UN\otimes\rho^*)$  expresses the fact that
$D_p\times(D_0\times D_q)\to D_p\times D_q\to(D_p\times D_0)\times D_q$
can be also obtained by applying the coassociator. Second, the pentagon
identity
  \be  (\cop\otimes\id\otimes\id)(\coa) \cdot (\id\otimes\id\otimes\cop)(\coa)
=
  (\coa\otimes\UN)\cdot(\id\otimes\cop\otimes\id)(\coa) \cdot(\UN\otimes\coa)
  \labl{pen}
expresses the equality of the two possible ways to get from
$((D_p\times D_q)\times D_r)\times D_s$   to
$D_p\times(D_q \times(D_r\times D_s))$.

Up to now we have that \H\ is a weak quasi-bialgebra while the
category $\C_H$ of its representations is a monoidal category.
To define contragredient representations (making $\C_H$
rigid) one passes from a bialgebra to a Hopf algebra --
 \h\ is endowed with an antipode, a linear *-antiautomorphism
  $\apo\!:\ \h\to\h$,
and non-zero elements $l,r\in\H$ such that
$  a^{(1)}\cdot l\cdot \apo(a^{(2)})=l\cdot\cou(a)$,
 $ \apo(a^{(1)})\cdot r\cdot a^{(2)} =\cou(a)\cdot r$\,
for all $a\in\h$. (We use Sweedler type notation, i.e.\ write
$\cop(a)=a^{(1)}\otimes a^{(2)}$ etc.)
Again there are compatibility constraints, namely the
  square identities
$  \apo(\lambda)\cdot\apo(\coa_1)\cdot r\cdot\coa_2\cdot l\cdot
  \apo(\coa_3)\cdot\apo(\rho^*) =\UN= \lambda^*\cdot\coa_1^*\cdot l\cdot
  \apo(\coa_2^*)\cdot r\cdot\coa_3^*\cdot\rho$\,.

\bull
Finally we want $\C_H$ to be braided, i.e.,
the coproduct is quasi-co\-com\-mu\-ta\-tive. Thus there is an element
$\coc\in\H\otimes\H$ such that
 $\copp(a)\cdot\coc= \coc\cdot\cop(a)$  for all $a\in\h$.
Here $\copp\equiv\tau\circ\cop$ ($\tau$ permutes the factors).
Again because of the weakness property one requires only that
$\coc$ is a partial isometry.
Compatibility requires the hexagon identities
  \begin{eqnarray} &&\coa_{231}\cdot(\cop\otimes\id)(\coc)\cdot\coa_{123}
  =\coc_{13}\cdot\coa_{132}\cdot\coc_{23} \,, \label{hex1} \\[1.5 mm]
  &&\coa_{312}^*\cdot(\id\otimes\cop)(\coc)\cdot\coa_{123}^*=\coc_{13}\cdot
  \coa_{213}^*\cdot\coc_{12}\, .  \label{hex2} \end{eqnarray}
Here we use the notation $\coa\equiv\coa_{123}=
\sum_i \coa_{1,i}\otimes\coa_{2,i}\otimes\coa_{3,i}=
\coa_1\otimes\coa_2\otimes\coa_3$; similarly, $\coa_{231}:=\coa_2
\otimes\coa_3 \otimes\coa_1$, $\coc_{13}:=\coc_1\otimes\UN\otimes\coc_2$, etc.

\smallskip\noindent
{\bf 5. Gauge freedom.}
We should not distinguish between two RHAs that possess the same category
of representations. This leads one to consider the following gauge
transformations, or twistings, of $H$. Set
$  {\cal U}_2^{}:=\{U\in\H\otimes\H \mid UU^*=\UN\otimes\UN\} $.
For $U\in{\cal U}_2$ define the twisted RHA to be given by
$(\H,\cou,\cop_U,\rho_U,\lambda_U,\coc_U,\coa_U,S,l_U,r_U)$, where
$\cop_U(a):=U\cop(a)U^*$,
$\rho_U=\cou(U_1)U_2\rho$, $\lambda_U=U_1\cou(U_2)\lambda$,
$l_U=U_1 l\apo(U_2)$, $r_U^{}=\apo(U_1^*)r U_2^*$,
$\coc_U=U_{21}\coc U_{12}^*$ and
$  \coa_U=U_{12}^{}\,[(\cop\otimes\id)(U)]\,\coa\, [(\id\otimes\cop)(U^*)]\,
  U_{23}^*$.

\smallskip\noindent
{\bf 6. Amplimorphisms, monodromy matrix, statistics parameters.}
An amplimorphism of $\H$ is a *-algebra monomorphism from \h\ to $M_n(\H)$
(the $n\times n$ matrices with entries in $H$).
One can define subobjects, direct sums, and an associative product
$(\mu\times\nu)^{i_1j_1,i_2j_2}(a):=\mu^{i_1i_2}(\nu^{j_1j_2}(a))$
of amplimorphisms.
 Any non-zero representation $D$ of \H\ of dimension $m$
defines a {\em special amplimorphism\/} $\mu_D\colon H\to M_m(H)$ via
  \be  \mu^{}_D:=(\id\otimes  D)\circ\cop \,. \ee
The braiding of amplimorphisms is described by the statistics operators.
 For special amplimorphisms $\mu_1,\, \mu_2$ corresponding to
representations $D_1$ and $D_2$, the statistics operator
  $ \stopt(\mu_1;\mu_2)=[(\id\otimes D_2\otimes D_1)](\coa\cdot
  \tau_{23} \coc_{23} \cdot \coa^*)$
($\tau$ interchanges the tensor product factors of the underlying
\rep\ spaces)
is an intertwiner between $\mu_2\times\mu_1$ and $\mu_1\times\mu_2$.
 For special amplimorphisms with non-zero $D$, there is
a partial isometry $P_\mu \in(\mu_{\bar D}\times\mu_D\vert\id)$
given by
 $ P_\mu^{ij,\cdot}= ({\rm tr}\, D(rr^*))^{-1/2}\,\coa_1
  \cdot D^{ji}(\coa_3 r^*\apo(\coa_2))$, $i,j=1,\ldots,{\rm dim}\,D$.
A {\em standard left inverse\/}
$\Phi_\mu\colon\, M_m(\H)\to \H$ of a special amplimorphism
$\mu\colon\, \H\to M_m(\H)$ is then  defined as
  $ \Phi_\mu(A)=P^*_\mu\cdot\bar \mu_D(A)\cdot P_\mu$ for all $A\in M_m(\H)$.
$\Phi_\mu$ is a unit preserving positive linear map satisfying
$\Phi_\mu(\mu(a)\cdot B\cdot\mu(c))=a\cdot\Phi_\mu(B)\cdot c$
for all $a,c\in\H$ and all $B\in M_m(\H)$.

The {\em statistics parameter matrix\/} $\Lambda_\mu\in M_m(\H)$ and the
{\em statistics parameter\/} $\lambda_\mu\in\H$ of an
amplimorphism $\mu_D\colon\, \H\to M_m(\H)$ are defined as
  $  \Lambda_\mu=\Phi_\mu(\stopt_\mu)$,
 $ \lambda_\mu=\Phi_\mu(\Lambda_\mu) $,
where $\stopt_\mu\equiv\stopt(\mu;\mu)$.
The statistics parameter $\lambda_\mu$ is an element of the
center of \H\ and depends only on the equivalence class of $\mu$. For an
irreducible $\mu_r=\mu^{}_{D_r}$, $ r\in\hh$, $\Lambda_{\mu_r}$ takes the form
  $  \Lambda_{\mu_r}^{}={\omega_r\over d_r}\cdot\mu_r(\UN)$, and hence
  $ \lambda_r={\omega_r\over d_r}\cdot \UN $.

Now we can explain the last part of the definition of a RHA,
namely the invertibility of the monodromy matrix
 $Y\in M_{\vert\hh\vert}(\H)$. $Y$ is defined by
  $  Y_{rs}:=d_rd_s\cdot\Phi_r\Phi_s(\stop(\nu_r;\nu_s)\cdot
  \stopt(\nu_s;\nu_r))$, $\,r,s\in\hh $.
One can show that $Y_{rs}=y_{rs}\cdot\UN$ with $y_{rs}\in\Co$.
As in the case  of rational field theory one can show that
$Y$ is invertible iff   $\,|\sigmA|^2=\sumh r d_r^2\,$
where  $ \sigmA := \sumh r d_r^2\,\omega_r^{-1} $
(if the monodromy matrix is degenerate,
then the algebra \h\ is said to be degenerate, too). Moreover in
the nondegenerate case the matrices
  \be  \MS_{rs}:={1\over \vert\sigmA\vert}\cdot y_{rs},\qquad
  \MT_{rs}:=\left({\sigmA\over\vert\sigmA\vert}\right)^{1/3}\cdot
  \delta_{rs}\,\omega_r  \label{mod} \ee
provide a unitary representation of the modular group, and
\be
    c = \frac{4\ii}\pi \log \frac \sigma{|\sigma|} \in [0,8)
\ee
plays the role of the `central charge' of \h, which should be
equal (mod 8) to the Virasoro central charge of any conformal field
theory model that has \h\ as its quantum symmetry.

The statistics parameters and the monodromy matrix are
independent of the gauge freedom described above, while the statistics
operators are invariant up to unitary equivalence.

\smallskip\noindent
{\bf 7. Polynomial equations.}
Now we consider in more detail the structure of RHAs.
The product of two irreducible representations in general is reducible,
i.e., we have $D_p\times D_q = \bigoplus_r N^r_{pq} D_r$ with
$N^r_{pq}=\dim(D_p\times D_q|D_r)$.
In terms of the representation spaces one has
$V_p\otimes V_q \supseteq\bigoplus_r (D_p\times D_q|D_r) \otimes V_r$, and
hence
   \be  n_p n_q \ge \sum_r N^r_{pq} n_r  \,. \labl,
This becomes an equality only if $\cop(\bfe)=\bfe\otimes\bfe$.
A function $\hh\ni r\mapsto n_r\in{\bf N}$ satisfying the
inequality \erf, is called a weak dimension function (finding such $n_r$
is easy, and in fact there are infinitely many solutions).

A basis in the intertwiner space $(D_{p_1}\times D_{p_2}|D_r)$ is given by
\CGC s $\cgc{p_1}{p_2}r{i_1}{i_2}k\alpha$, where \,$p_j,r\in\hh$,
$\alpha\in\{1,2,...\,,N_{p_1p_2}^r\}$, and $i_j\in\{1,2,...\,,n_{p_j}\}$,
etc. They contain the same information as the coproduct; indeed
on matrix units $e^{i,i^*}_r\in M_t\subseteq H$, the coproduct \cop\ acts as
  $ \cop(e^{i,i^*}_r) = \sum \cgc {p_1}{p_2}r{k_1}{k_2}i\alpha
   \cgcs {p_1}{p_2}r{k^*_1}{k^*_2}{i^*}\alpha
  \,e_{p_1p_2}^{k_1k_2,k^*_1k^*_2} $
with $p_j\in\hh$, $\,k^{(*)}_j\in\{1,\dots,n_{p_j}\}$,
$\alpha\in\{1,\dots, N^r_{p_1p_2}\}$ (we denote by
  $ \e {p_1p_2...p_m} {i_1i_2...i_m} {j_1j_2...j_m}
  \equiv \epij 1\otimes \epij 2\otimes\ldots\otimes \epij m $
 the matrix units of $\h^{\otimes m}$).  The fact that \cop\ is an
algebra homomorphism implies the orthogonality property
  $ \sumi ir \sumi js \cgcs rsuijk\alpha \cgc rsvijl\beta
  = \delta_{uv}\delta_{kl}\delta_{\alpha\beta} $.

The most general form  of the coassociator reads
  \be   \coa = \sum \F{pqr}\alpha u\beta \gamma v\delta t
  \cgc pqu i{i'}l\alpha \cgc urt l{i''}k\beta
  \cgcs qrv {j'}{j''}m\gamma \cgcs pvtjmk\delta    \,
  \e{pqr}{ii'i''}{jj'j''} \, \labl{coa}
with $\Fpqr\in\complex$ and the sum being over all labels that appear
in the expression.
{}From the requirements that $\coa$ is a partial isometry it follows
that for fixed $p$, $q$,
$r$ and $t$ for which $\FF pqr\cdot\cdot t$ are non-vanishing, \Fpqr\
is a unitary matrix in the (multi-)indices $(\alpha,u,\beta)$ and
$(\gamma,v,\delta)$, i.e.
  \be  \sumh w \sumn \mu qrw \sumn \nu pwt \F{pqr}\alpha u\beta\mu w\nu t
  \Fb{pqr}\gamma v\delta\mu w\nu t=\delta_{\alpha\gamma}\delta_{\beta\delta}
  \delta_{uv}  \labl u
for $N_{pq}^{\ \,u}N_{ur}^{\ \,t}>0$ and $\alpha\in\{1,2,...\,,N_{pq}
^{\ \,u}\}$, $\beta\in\{1,2,...\,,N_{ur}^{\ \,t}\}$.

The cocommutator \coc\ intertwines \cop\ and
\copp; thus it can be written as
  \be  \coc = \sum
  \R rst\alpha\beta
  \cgc srt{i'}ik\alpha \cgcs rstj{j'}k\beta \, \e{rs}{ii'}{jj'} \, \labl{coc}
with $\R pqt\alpha\beta\in\complex$ and again a sum over `everything'.
Because \coc\ is a partial isometry, for fixed $p$, $q$ and $t$,
\R pqt\alpha\beta\ is (if non-vanishing) a unitary matrix in the
indices $\alpha$ and $\beta$.

Let us now rewrite the pentagon identity \erf{pen} in terms of the
$F$-matrices that are defined by \erf{coa}.
Using the orthogonality of the \CGC s, we obtain
  \be  \sumn\sigma uvt \Fa{pqv}\sigma u\alpha \mu y\beta  t
                 \cdt \Fa{urs}\nu x\delta \sigma v\gamma t
  = \sumh w \sumn\kappa wsy \sumn\lambda pwx \sumn\eta qrw
                       \Fa{pqr}\delta u\alpha \lambda w\eta x
                 \cdt \Fa{pws}\nu x\lambda \mu y\kappa t
                 \cdt \Fa{qrs}\kappa w \eta \beta v\gamma y \,. \labl{penf}
Similarly, with \erf{coc} the hexagon identities read
  \be \begin{array}{l}
  {\dstyle \sumh u \sumn{\delta,\lambda}urt \sumn\kappa pqu }\,
  \F{rpq}\beta w\gamma\kappa u\delta t \cdt \R urt\delta\lambda \cdt
  \F{pqr}\kappa u\lambda\mu v\nu t= {\dstyle\sumn\alpha prw\sumn\delta qrv }\,
  \R prw\beta\alpha \cdt \F{prq}\alpha w\gamma\delta v\nu t \cdt
  \R qrv\delta\mu \,, \\{} \\[-2.1 mm]
  {\dstyle \sumh u \sumn{\delta,\lambda}urt \sumn\kappa pqu }\,
  \Fb{rpq}\beta w\gamma\kappa u\delta t \cdt \R rut\lambda\delta \cdt
  \Fb{pqr}\kappa u\lambda\mu v\nu t= {\dstyle\sumn\alpha prw\sumn\delta qrv}\,
  \R rpw\alpha\beta \cdt \Fb{prq}\alpha w\gamma\delta v\nu t \cdt
  \R rqv\mu\delta \,.  \end{array} \labl{hexf}

\smallskip\noindent
{\bf 8.  Outlook.}
The essential information of $\C_\A$ consists of the fusion rules
 $\{N_{pq}^r\}$ and the fusing and braiding matrices
 $\{F^{(pqr)_{\scriptstyle t}}_{}, R_{}^{(pq)_{\scriptstyle r}}\}$
(i.e., the category can be
reconstructed from this information \cite{Moore-Seiberg}).
As we see from the results above, given these data together with
a weak dimension function $\{n_r\}$,
one can construct a RHA \h\ such that its category
$\C_H$ will be equivalent to $\C_\A$ (see \cite{SV2,K,S,H,M} for details).
Thus with the correct definition the \DR reconstruction
becomes `almost a tautology' for rational theories.

Consider now the character rings. The character ring
$[\C_\A]$ is a MFA characterized by the fusion rules $\{N_{pq}^r\}$
and the statistical weights $\{\omega_r\}$.
(The dimensions $\{d_r\}$ are determined by the fusion rules,
while from the statistical phases and the fusion rules one easily
recovers the $\MS$ and $\MT$ matrices of the modular group.)
Obviously also every RHA (more precisely, every equivalence class [RHA]
of RHAs modulo twisting and modulo the choice of weak dimension function)
defines a MFA. An interesting but highly nontrivial
problem is to analyze the map $$ [{\rm RHA}] \longrightarrow {\rm MFA} \,.  $$

Is this map one to one? In other words, are the statistical
phases (and, of course, the fusion rules)
enough to distinguish between different [RHA], or
equivalently between different categories $\C_\A$\,? Most likely the answer to
this question is yes.

Is this map onto? This is a much  more difficult question. Ideally one would
like, given $\{N_{pq}^r\}$ and  $\{\omega_r\}$, to have an ansatz for
the fusing and braiding matrices
$\{F^{(pqr)_{\scriptstyle t}}_{}, R_{}^{(pq)_{\scriptstyle r}}\}$ in terms of
statistical phases and dimensions that solve the polynomial equations.
A positive answer to this question will mean
that one can reconstruct the braid group representation
of a conformal model from its modular properties.

We have explored this problem only in a tiny corner of the space of
all MFAs. All possible MFAs of dimension $\le 3$ have been classified.
Taking only the fusion rule data  $\{N_{pq}^r\}$ of these MFAs one
can find the general solution to the corresponding polynomial equations.
Referring to \cite{FGV} for the details, here we only mention that
for $|\hh|\le 3$ the map in question is an isomorphism.

A positive answer to the posed question will mean that
the classification of MFAs (a formidable problem in itself)
will lead to a classification of the Moore\hy Seiberg categories
$\C_\A$. The next natural question will be how far is a
classification of MFAs from a classification of, say,
rational conformal field theories. It is known that there
are different models (having for example different Virasoro
central charges) that share the same braiding properties
and hence Moore\hy Seiberg categories. Hence one can ask
what additional information is necessary to distinguish
between them. In particular, are the conformal weights
(i.e., not just their fractional parts which are the statistical weights)
already sufficient?

\vskip1em \medskip\noindent\small
{\bf Acknowledgement.} A.G.\ would like to thank the
organizers of the Karpacz Winter School '94 for being invited to
participate and for the opportunity to present this talk.

   \def\nupb {Nucl.\wb Phys.\wb B}
   \def\bams {Bull.\wb Amer.\wb Math.\wb Soc.}
   \def\comp {Commun.\wb Math.\wb Phys.}
   \def\inma  {Invent.\wb math.}
   \def\jams  {J.\wb Amer.\wb Math.\wb Soc.}
   \def\jgap  {J.\wb Geom.\wB and\wB Phys.}
   \def\joal  {J.\wB Al\-ge\-bra}
   \def\jodg  {J.\wb Diff.\wb Geom.}
   \def\jofa  {J.\wb Funct.\wb Anal.}
   \def\jopa  {J.\wb Phys.\ A}
   \def\jomp  {J.\wb Math.\wb Phys.}
   \def\lemp  {Lett.\wb Math.\wb Phys.}
   \def\lenc  {Lett.\wB Nuovo\wB Cim.}
   \def\leni  {Lenin\-grad\wB Math.\wb J.}
   \def\maan  {Math.\wb Annal.}
   \def\mams  {Memoirs\wB Amer.\wb Math.\wb Soc.}
   \newcommand\mapx{in: {\sl Mathematical Physics X},
              {K.\ Schm\"udgen, ed.}, \SV, \Be {1992}}
   \def\mpla  {Mod.\wb Phys.\wb Lett.\ A}
   \newcommand{\mqft}[2] {\inBO{Modern Quantum Field Theory}
              {S.\ Das, A.\ Dhar, S.\ Mukhi, A.\ Raina, and A.\ Sen, eds.}
              \WS\Si{1991} {{#1}}{{#2}}}
   \def\npbf  {Nucl.\wb Phys.\ B\vypf}
   \def\npbp  {Nucl.\wb Phys.\ B (Proc.\wb Suppl.)}
   \newcommand{\nspq} {in: 1991 Carg\`ese Lectures on
              {\sl New Symmetry Principles in Quantum Field Theory},
              {J.\ Fr\"ohlich et al., eds.}, Plenum Press, \NY {1992}}
  \def\Be     {{Berlin }}
   \def\Ca     {{Cambridge}}
   \def\NY     {{New York }}
   \def\pR     {{Princeton}}
   \def\PR     {{Providence}}
   \def\Si     {{Singapore }}
 \def\SV     {{Sprin\-ger Verlag}}
   \def\WI     {{Wiley Interscience}}
   \def\WS     {{World Scientific}}
 \def\phlb  {Phys.\wb Lett.\ B}
 \def\slnm  {Sprin\-ger\wB Lect.\wb Notes\wB in\wB Math.}
  \newcommand{\Suse} {in: {\sl The Algebraic Theory of Superselection
              Sectors.\ Introduction and Recent Results}, {D.\ Kastler,
              ed.}, \WS, \Si {1990}}

 \end{document} \bye